\documentclass[a4paper,11pt]{article}
\usepackage{pos}

\title{
\vskip-3cm{\baselineskip14pt
    \begin{flushright}
     \normalsize \normalfont{CERN-TH-2026-152}
    \end{flushright}} \vskip2.5cm
Analytic two-loop electroweak corrections at high energies}

\author*[a]{Hantian Zhang}

\affiliation[a]{Theoretical Physics Department, CERN, \\
1211 Geneva 23, Switzerland}

\emailAdd{hantian.zhang@cern.ch}

\abstract{The high-energy behaviour of electroweak scattering amplitudes is of theoretical and phenomenological interest. In these proceedings, we summarize recent progress in analytic high-energy calculations for two-loop four-point electroweak amplitudes in the full Standard Model. As a representative application, we discuss the electroweak corrections to Higgs boson pair production, where rich structures of logarithmic and power corrections appear and sizeable effects are found in the high-energy region.}

\FullConference{Loops and Legs in Quantum Field Theory (LL2026)\\
12-17, April, 2026\\
Bayreuth, Germany\\}

\begin{document}
\maketitle

\section{Introduction}

There has been a long history of studying the high-energy behaviour of
scattering amplitudes in quantum field theory, starting from the classic works
of Sudakov and Weinberg~\cite{Sudakov:1954sw,Weinberg:1959nj}. In recent
years, electroweak scattering amplitudes in this limit have drawn renewed attention~\cite{Davies:2022ram,Davies:2025wke,Davies:2026wbx},
motivated by the need for high-precision predictions at the Large Hadron
Collider (LHC) and by the development of new analytic methods for multi-loop
calculations.

The state of the art has advanced in several directions. In QCD and QED,
high-energy expansions of massive two-loop amplitudes have been studied for a
number of phenomenologically important processes~\cite{Melnikov:2016qoc,Kudashkin:2017skd,Liu:2017vkm,Davies:2018ood,Davies:2018qvx,Mishima:2018olh,Catani:2022mfv,Buonocore:2023ljm,Wang:2023qbf,Wang:2024pmv,Lee:2024dbp,Lee:2024jvd,Jaskiewicz:2024xkd,Hu:2025aeo,Hu:2025hfc,Wang:2025kpk,Delto:2025epy}. Electroweak two-loop amplitudes in the
full Standard Model~(SM) are more involved because several mass scales appear simultaneously in
the Feynman diagrams. In the literature, logarithmic electroweak corrections at the leading power have been intensively studied~\cite{Fadin:1999bq,Hori:2000tm,Melles:2001ye,Kuhn:2001hz,Beenakker:2001kf,Denner:2003wi,Denner:2004iz,Pozzorini:2004rm,Jantzen:2005az,Denner:2006jr}, and have been recently implemented in \texttt{OpenLoops}~\cite{Lindert:2026ief,Buccioni:2019sur}.
Beyond logarithmic and leading-power approximations, the complete high-energy
expansion of two-loop four-point electroweak amplitudes in the full SM has only
been obtained recently in Ref.~\cite{Davies:2026wbx} for Higgs boson pair
production $gg \to HH$.

In these proceedings, we summarize the analytic roadmap for two-loop electroweak amplitudes at high energies. We first discuss the high-energy expansion strategy for electroweak
amplitudes and the \texttt{AsyInt}~\cite{Zhang:2024fcu,Zhang:2025ysy} approach to master integrals. We then
describe its application to the top-quark-induced next-to-leading order
electroweak~(NLO EW) corrections to $  g g \to H H$,
and show their sizable impacts in the high-energy region.

\section{High-energy expansion of electroweak amplitudes}

The high-energy limit of four-point electroweak amplitudes considered in the Feynman gauge is characterized by
\begin{equation}
  s, |t| \gg m_t^2, m_W^2, m_Z^2, m_H^2 .
  \label{eq:he_hierarchy}
\end{equation}
In this limit, the SM masses are treated as small parameters
compared with the Mandelstam variables. The dependence on the masses is explicitly kept in the
asymptotic high-energy expansion.
These masses appearing in the loop
diagrams are further expanded around an electroweak central scale, which can be conveniently chosen as the top-quark mass
\begin{equation}
  \delta_X = 1 - \frac{m_X}{m_{\rm EW}},
  \quad X = H,W,Z  \; \text{ and } \; m_{\rm EW}  = m_t.
  \label{eq:delta_def}
\end{equation}
Since the limiting behaviour of Eq.~\eqref{eq:he_hierarchy} is characterized by the power-log series in small masses, the mass-difference $\delta_X$ expansion becomes a Taylor expansion.
For $gg \to HH$, the calculation can be further simplified through a small external-Higgs-mass expansion, which is again a Taylor expansion.
Note that this small external-mass expansion is in principle not a necessary step, but can reduce the complexity of both amplitudes and integral calculations.
After these expansions and subsequent IBP reductions, the resulting master integrals $\vec{I}$ depend on the variables $s$, $t$ and $m_{\rm EW}$.

In the high-energy limit, a convenient way to proceed is to use
differential equations with respect to the mass parameter
\begin{equation}
  \frac{\partial}{\partial m_{\rm EW}^2} \vec{I}
  =
  M(s,t,m_{\rm EW}^2,\epsilon) \vec{I} ,
  \label{eq:de_mi}
\end{equation}
and plug in the power-logarithmic series ansatz for each master integral
\begin{equation}
   I_n
  =
  \sum_{i,j,k}
  C^{(n)}_{ijk}(s,t)\,
  \epsilon^i \,
  m_{\rm EW}^j \,
  \log^k(m_{\rm EW}^2).
  \label{eq:power_log_ansatz}
\end{equation}
The resulting linear system of equations from the differential equations can be solved in terms of unknown boundary
functions $C_{ijk}^{(n)}(s,t)$.
These functions can be extracted from the boundary master integrals, which are calculated
with \texttt{AsyInt}~\cite{Zhang:2024fcu} in a direct parametric integration approach incorporating the method of regions~\cite{Beneke:1997zp} .
The most complicated fully massive EW non-planar master integrals are computed in Ref.~\cite{Davies:2025wke}, for which we also employed the $t$-differential equations approach for the expanded master integrals as in Eq.~\eqref{eq:power_log_ansatz}, in order to compute the boundary functions at higher orders in $\epsilon$ and $m_{\rm EW}$.
The final analytic results of the master integrals are expressed in terms of Harmonic Polylogarithms and special transcendental constants.
For technical details regarding master integral calculations, we refer to Refs.~\cite{Zhang:2024fcu,Davies:2025wke}.

\section{Application to Higgs boson pair production}

\begin{figure}[t]
\centering
    \includegraphics[width=0.85\textwidth]{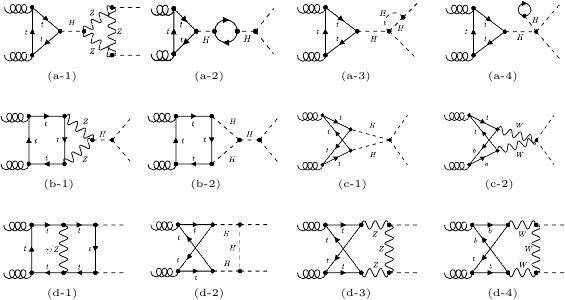}
  \caption{Sample Feynman diagrams contributing to
  the NLO EW corrections to $gg\to HH$.}
   \label{fig:ew_diagrams}
\end{figure}

We now discuss the application to the top-quark-induced electroweak corrections
to $gg\to HH$. The process is loop-induced at
LO and the NLO EW corrections require
two-loop amplitudes, whose sample diagrams are shown in Fig.~\ref{fig:ew_diagrams}.
In recent years, substantial efforts have been devoted to the computation of these contributions using different methods~\cite{Davies:2022ram,Muhlleitner:2022ijf,Davies:2023npk,Bi:2023bnq,Zhang:2024rix,Li:2024iio,Heinrich:2024dnz,Davies:2025wke,Bonetti:2025vfd,Bhattacharya:2025egw,Davies:2026wbx}.

The amplitude is decomposed into two Lorentz structures with scalar form
factors. Following the conventions of Ref.~\cite{Davies:2026wbx}, one writes
\begin{equation}
  {\cal M}^{\mu\nu}
  =
  F_1\,T_1^{\mu\nu}
  +
  F_2\,T_2^{\mu\nu} ,
  \label{eq:amp_decomp}
\end{equation}
and the form factors are
expanded perturbatively as
\begin{equation}
  F
  =
  F^{(0)}
  +
  \frac{\alpha_s}{\pi} F^{(1,0)}
  +
  \frac{\alpha}{\pi} F^{(0,1)}
  + \cdots .
  \label{eq:pert_expansion}
\end{equation}
The term $F^{(0,1)}$ denotes the electroweak form factors considered here.

The calculation is performed in several expansion steps, as described above. First, the masses of
the internally propagating particles are expanded around $m_{\rm EW} = m_t$ in the parameters
$\delta_X$ defined in Eq.~\eqref{eq:delta_def}.
Second, the final-state Higgs
boson mass is expanded at the integrand level. The resulting scalar integrals
are then reduced to master integrals and computed in the high-energy limit with \texttt{AsyInt} and differential equations (see Refs.~\cite{Davies:2018qvx,Mishima:2018olh,Zhang:2024fcu,Davies:2025wke}).

The renormalization is performed in the on-shell scheme for the input
parameters $e$, $m_W$, $m_Z$, $m_t$ and $m_H$. The result is then transformed to
the $G_\mu$ scheme. Tadpole contributions are included consistently, following
the \textit{Fleischer--Jegerlehner} tadpole scheme. 
For details of the renormalization procedure including the treatment of gauge-fixing parameters in the general $R_\xi$ gauge, we refer to Refs.~\cite{Davies:2023npk,Zhang:2024rix}.

The generic structure of the high-energy expansion of the form factors is
\begin{equation}
  F =
  \sum
  C_{i j n}^{ k_W k_Z k_H}(s,t)
  \, m_t^i \, 
  \log^j(m_t^2) \,
  \left(m_H^{\rm ext}\right)^{2n}
  \delta_W^{k_W}
  \delta_Z^{k_Z}
  \delta_H^{k_H} \,,
  \label{eq::generic_expansion}
\end{equation}
where the $m_H^{\rm ext}$ denotes the external Higgs mass. 
In practice, we compute deep high-energy expansions up to the orders $m_t^{108}$, $(m_{H}^{\rm ext})^4$ and $\delta_X^4$ are obtained and used to construct
Pad\'e-improved approximations, in order to cover the phase space region down to fairly low values of Higgs transverse momentum $p_T$.

The analytic result also allows one to inspect the logarithmic structure of the
form factors. For example, in the $m_t \to 0$ and $m_H^{\rm ext} = 0$ limit, the
leading logarithmic terms of the box form factors take the form
\newcommand{\vph}{\vphantom{\Big\{}}
\newcommand{\lms}{l_{ms}}
\newcommand{\lts}{l_{ts}}
\newcommand{\lots}{l_{1ts}}
\begin{align}
    F_{\rm box1}^{(0,1)} & = 
    \frac{m_W^4}{s^2  c_{\rm w}^6 s_{\rm w}^2} \Bigg[ \, l_{ms}^2 \,
    \frac{36 c_{\rm w}^6+32 c_{\rm w}^4-40 c_{\rm w}^2+17 }{36}  
    \Big(l_{1ts}^2+l_{ts}^2 -2 l_{1ts} l_{ts} +\pi ^2-4\Big)
    \Bigg] + \cdots,
    \label{eq::Fbox12l}\\[5pt]
    F_{\rm box2}^{(0,1)} & = 
    \frac{m_W^4}{s^2  c_{\rm w}^6 s_{\rm w}^2} \Bigg[ \, l_{ms}^3 \, \frac{36 c_{\rm w}^6+32 c_{\rm w}^4-40 c_{\rm w}^2+17}{27}   +\,  l_{ms}^2 \, \bigg(
     -\frac{2}{9}\, \delta_Z \, \left(32 c_{\rm w}^4-40 c_{\rm w}^2+17\right) \nonumber \\
     &\quad -8 \, \delta_W \, c_{\rm w}^6 +  
    \frac{36 c_{\rm w}^6+32 c_{\rm w}^4-40 c_{\rm w}^2+17 }{36 t (s+t)} \Big(l_{1ts}^2 (s+t)^2-2 \, l_{1ts} \, t (s+t)-2  \,l_{ts} \, t (s+t) \nonumber \\
    &\quad +\, l_{ts}^2 \, t^2 +\pi ^2 \left(s^2+2 s t+2 t^2\right)\Big)\bigg)
    \Bigg] + \cdots,
    \label{eq::Fbox22l}
\end{align}
where  $c_{\rm w}\equiv\cos(\theta_{\rm w})$
is the cosine of the weak mixing angle and
\begin{eqnarray}
  \lms  &=& \log \left(\frac{m_t^2}{s}\right) +i\pi\,,\quad
  \lts  \,\,=\,\, \log\left(-\frac{t}{s}\right) +i\pi\,,\quad
  \lots \,\,=\,\, \log\left(1+\frac{t}{s}\right) +i\pi\,.
\end{eqnarray}
Note that the counterterms do
not contribute to the results 
shown in Eqs.~(\ref{eq::Fbox12l}) and~(\ref{eq::Fbox22l});
they come exclusively from the two-loop diagrams. 
We further note that  $gg \to HH$ at NLO EW
contains several coupling structures, resulting in rich structures of logarithmic and
power corrections.
For further analytic expressions we
refer to the ancillary files of Ref.~\cite{Davies:2026wbx}.

\begin{figure}[t]
\centering
\begin{tabular}{c}
    \includegraphics[width=.6\textwidth]{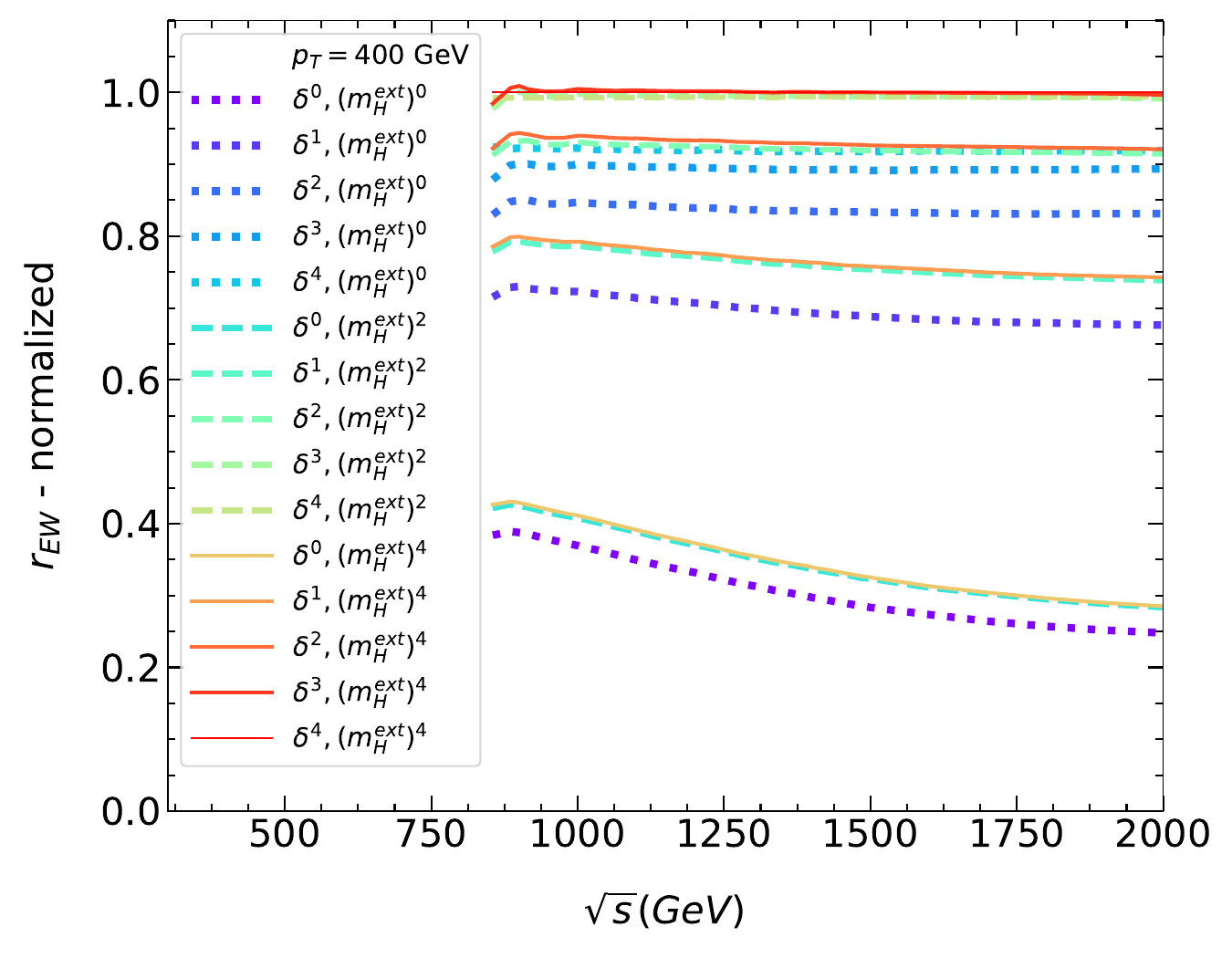}
\end{tabular}
  \caption{\label{fig::FF_he_ME_ra}
   $r_{\rm EW}$ for various $\delta$ and $m_H$ expansion terms normalized to
   highest available approximation for  $p_T=400$~GeV.
   The truncation uncertainty in $\delta_X$ and $m_H^{\rm ext}$ expansions is estimated conservatively to $\pm1\%$.
   }
\end{figure}

\begin{figure}[t]
\centering
    \includegraphics[width=.6\textwidth]{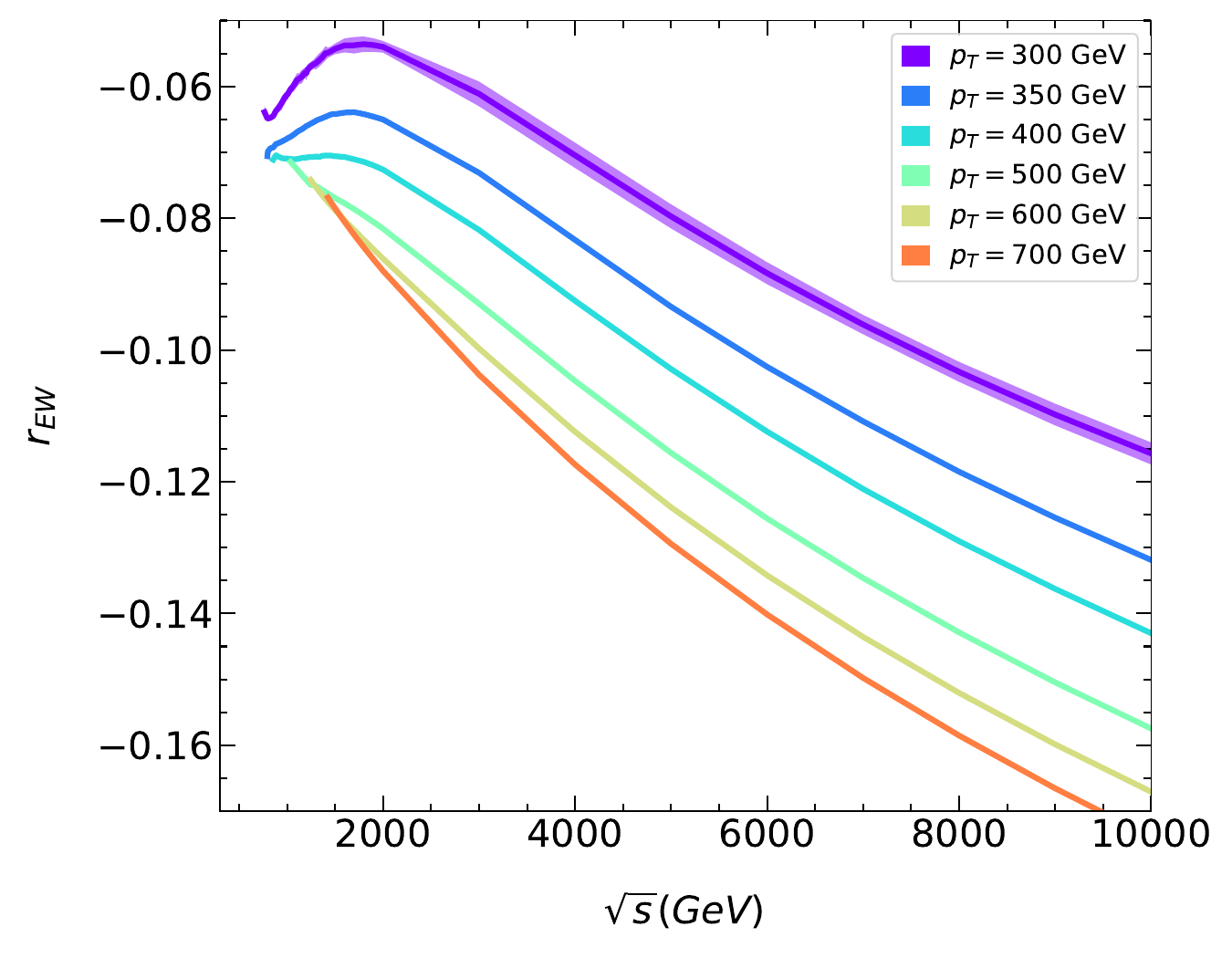}
  \caption{\label{fig::FF_he_ratio}
   $r_{\rm EW}$ for various values of $p_T$ as a function of $\sqrt{s}$.
   The plots show the same data for different ranges of $\sqrt{s}$ on the $x$ axis. The uncertainty band (only visible for $p_T = 300$~GeV) is obtained from the Pad\'{e} approximation.}
\end{figure}

For numerical applications, the correction can be quantified through the ratio
\begin{equation}
  r_{\rm EW}
  =
  \frac{\alpha}{\pi}
  \frac{\mathcal{U}^{(0,1)}}{\mathcal{U}^{(0)}} \,,
  \label{eq:rew}
\end{equation}
where the  squared matrix element is expanded as $|\mathcal{M}|^2 = \bar{X}_0 \big( \, \mathcal{U}^{(0)} + \frac{\alpha_s}{\pi}\, \mathcal{U}^{(1,0)} + \frac{\alpha}{\pi} \, \mathcal{U}^{(0,1)} \big)$ with an overall prefactor $\bar{X}_0$.
The high-energy expansion shows good convergence in a broad region of phase space, as illustrated in Fig.~\ref{fig::FF_he_ME_ra} for $p_T = 400$~GeV.
The remaining uncertainty is dominated by the mass-difference expansions, and is conservatively estimated to be $\pm 1\%$ of the
two-loop electroweak correction.

In Fig.~\ref{fig::FF_he_ratio}, we show our results up to $\sqrt{s} = 10$~TeV.
The resulting top-quark-induced electroweak correction is negative and reaches
the order of $-10\%$ at high energies, supporting the observation from Ref.~\cite{Bi:2023bnq} for the hadronic invariant-mass or $p_T$ distribution.
The expansion remains reliable down to
transverse momenta of about $p_T=350\,{\rm GeV}$, providing an analytic result
for the boosted region of Higgs-boson pair production.

\section{Summary}

In these proceedings, we summarized recent progress towards analytic two-loop
electroweak corrections at high energies. The central technical ingredient is
the high-energy expansion of electroweak amplitudes and the calculation of massive two-loop four-point master integrals in this limit. This is achieved with the differential equation approach in combination with the \texttt{AsyInt} approach based on the direct parametric integrations and the method of regions.

As an application, we discussed the top-quark-induced electroweak corrections
to $gg\to HH$. The result is obtained as a deep high-energy expansion,
supplemented by mass-difference and external-Higgs-mass expansions. The
analytic form factors exhibit rich structures of logarithmic and power corrections, and
the corresponding electroweak correction is negative and of order $-10\%$ in
the high-energy region.
The expansion shows good
convergence and remains reliable down to transverse
momenta of about $p_T \ge 350\,{\rm GeV}$. These results provide analytic input for boosted
Higgs-pair production and illustrate the usefulness of high-energy expansions
for electroweak two-loop amplitudes.

\section*{Acknowledgment}
The author thanks Joshua Davies, Kay Sch\"onwald, Matthias Steinhauser for collaborations. This research is funded by the European Union under the Marie Sk{\l}odowska-Curie Actions (MSCA) grant 101202083 -- ``HINOVA''.

\small
\setlength{\bibsep}{4pt}
\bibliographystyle{JHEP}
\bibliography{reference}

\end{document}